\documentstyle[12pt,aasms]{article}

\begin{document}
\newcommand{\etal}{{\it et al.}}
\include{idl}

\title{Pulse Profiles, Accretion Column Dips and a Flare in GX 1+4 During a Faint State}
\author{A. B. Giles$^{1,2,3}$, D. K. Galloway$^{1,4}$, J. G. Greenhill$^{1}$, 
M. C. Storey$^{4}$, C. A. Wilson$^{5}$}
\affil{$^{1}$ School of Mathematics and Physics\\
              University of Tasmania\\
	      GPO Box 252-21, Hobart\\
              Tasmania 7001, Australia\\
       $^{2}$ {\it also} Laboratory for High Energy Astrophysics\\
              Mail Code 662, NASA Goddard Space Flight Center\\
              Greenbelt, MD 20771, USA\\ 
       $^{3}$ {\it also} Universities Space Research Association\\
       $^{4}$ Special Research Centre for Theoretical Astrophysics\\
              School of Physics, University of Sydney\\
	      Sydney, NSW 2006, Australia\\
       $^{5}$ Space Science Laboratory\\
              NASA Marshall Space Flight Center\\
              Huntsville, AL 35812, USA}


\begin{abstract}
The {\it Rossi X-ray Timing Explorer} (RXTE) spacecraft observed the X-ray 
pulsar GX 1+4 for a period of 34 hours on July 19/20 1996. The source 
faded from an intensity of $\sim20$ mcrab to a minimum of $\leq0.7$ 
mcrab and then partially recovered towards the end of the observation. This 
extended minimum lasted $\sim$40,000 seconds. Phase folded light curves 
at a barycentric rotation period of 124.36568 $\pm$ 0.00020 seconds show that 
near the center of the extended  minimum the source stopped pulsing 
in the traditional sense but retained a weak dip feature at the rotation 
period. Away from the extended minimum the dips are progressively 
narrower at higher energies and may be interpreted as obscurations or eclipses 
of the hot spot by the accretion column. The pulse profile changed from 
leading-edge bright before the extended minimum to trailing-edge bright 
after it. Data from the {\it Burst and Transient Source Experiment} 
(BATSE) show that a torque reversal occurred $<10$ days after 
our observation. Our data indicate that the observed rotation departs 
from a constant period with a $\dot P/P$ value of $\sim -1.5$\% per year 
at a 4.5$\sigma$ significance. We infer that we may have serendipitously 
obtained data, with high sensitivity and temporal resolution about the 
time of an accretion disk spin reversal. We also observed a rapid flare 
which had some precursor activity, close to the center of the extended 
minimum. 
\end{abstract}


\section{INTRODUCTION}
The binary X-ray pulsar GX 1+4 is unique in many respects.  It is the only 
known hard X-ray source in a symbiotic system. Its optical companion is 
V2116 Oph, a 19th magnitude M6 III giant.  It has the largest known rate 
of change of pulse period, with $\dot P/P$ as high as 2\% per year in 
the 1970's. The relationship between the spin period history and the 
luminosity is much more complex than is predicted by standard accretion 
theory (Ghosh \& Lamb 1979), with sustained periods observed where the 
spin-down rate was /it inversely /tm correlated with X-ray flux 
(Chakrabarty et al. 1997). Furthermore, GX 1+4 is thought to have a 
magnetic field of $\sim3$ $\times 10^{13}$ Gauss (Dotani et al. 1989; 
Greenhill et al. 1993; Cui 1997) which is amongst the strongest known 
for any object. 

The mean X-ray flux from GX 1+4 is variable on timescales of days to decades. 
In the decade following its discovery in 1971 (Lewin, Ricker \& McClintock 
1971) the flux was persistently high at $>100$ mcrab (McClintock \& 
Leventhal 1989). During the early 1980's the flux decreased by several 
orders of magnitude (to $<0.5$ mcrab on one occasion, Mukai 1988) and 
since then the source has normally been toward the lower end of its 
historical intensity range. There is considerable uncertainty about the 
source distance of 3 - 15 kpc (Chakrabarty \& Roche 1997) and hence about 
the X-ray luminosity. The very long-term flux variations are correlated with 
both spin period history and pulse profiles. During the 1970s when the 
pulsar was spinning up steadily the typical pulse profiles in most 
energy bands were broad, brighter on the trailing-edge and sometimes 
with a secondary minimum (see e.g. Doty, Hoffman \& Lewin 1981). With the 
change to a lower flux state in the 1980s the pulsar began steady spin-down 
(with brief returns to spin-up) and the observed pulse profiles were 
typically leading-edge bright (Greenhill, Galloway \& Storey 1998). 
The orbital period is unknown but is thought to be of the order of 
one year (Cutler et al. 1986).

\section{THE OBSERVATIONS}
The {\it Rossi X-ray Timing Explorer} (RXTE) observed GX 1+4 for a useable 
total of 51,360 seconds during a $\sim$34 hour period starting at 16:46 UT 
on 19th July 1996.  The observations were made in 23 sections with the 
interruptions being due to regular passages through the South Atlantic 
Anomaly (SAA) and Earth occultations of the source. Five additional short 
gaps were due to RXTE monitoring campaigns on other sources. The variation 
in intensity seen by the Proportional Counter Array (PCA) detectors 
(Zhang et al. 1993) is shown in the upper panel of Fig. 1. The data are 
shown for Proportional Counter Units (PCU's) 1, 2 \& 3 since PCU 4 was 
turned off after 0:03 UT on 20th July and PCU 5 after 2:32 UT on 20th 
July. These two detectors are occasionally commanded off to provide high 
voltage rest periods. Throughout this paper all analysis results and 
figures apply to the summation of PCU's 1, 2 \& 3 only, all the data have 
been background subtracted and all quoted UT values are barycentric times. 
The principal observing mode used was E\_250US\_128M\_0\_8S and so every 
detected X-ray photon was time tagged to 250$\mu$s and its energy was 
measured in 128 pulse height channels.

These RXTE observations were proposed with the particular intention of 
studying the high energy spectrum of GX1+4. However, the source turned 
out to be much weaker than expected in all energy bands so, in this paper 
we do not discuss any data from the High Energy X-ray Timing Experiment 
(HEXTE) on RXTE. Since the source was relatively faint the PCA background 
subtraction becomes critical. We have used the RXTE guest observers 
PCABACKEST software to estimate the various instrumental and orbital 
background contributions. The background model used was 
"sky VLE". Due to the proximity of GX 1+4 to the galactic plane 
we have also subtracted a 'galactic ridge' emission component determined 
from data in Valinia \& Marshall (1998). Their results were used to 
calculate integrated count rates in the direction of GX 1+4 for our light 
curve energy ranges of 2 - 7 keV \& 7 - 20 keV. The derived rates were 
also compared to the PCA count rates seen in the vicinity of GX 1+4 for 
the various slews to and from the source which were made during our 
observation. For the 2 - 7 keV range we have taken a value of 3.46 
counts s$^{-1}$ and for the 7 - 20 keV range 0.64 counts s$^{-1}$. 
We estimate that any remaining systematic error in our background 
subtraction is $\sim0.5$ counts s$^{-1}$. 

We do not discuss any spectral analysis results in this paper since our 
primary concern here is with timing issues. We do, however, see clear 
spectral variations with pulse phase and a full presentation of the 
phase resolved spectral analysis can be found in Galloway et al. (1999) 
which also discusses the spectral evolution through the flare reported 
here.

\section{TEMPORAL STUDIES}
During the observation the flux decreased to a minimum after about 20 hours 
and then began a gradual recovery towards its initial intensity. The light 
curve is shown in the upper panel of Fig. 1 and is plotted as a 4 second 
running average with a time resolution of one second. The Crab produces 
$\sim$7,800 counts s$^{-1}$ in 3 PCU's so the source flux during this 
observation ranges from $\leq0.7$ to $\sim20$ mcrab. We have divided the 
complete observation into the three intervals marked as 1 - 3 in Fig. 1. 
The $\sim$124 second rotation modulation creates a wide 
scatter of points and during the lowest intensity state in Fig. 1 (interval 
2) the minima appear to almost reach zero. For data which are not running 
mean averaged, as in Fig. 1, the one second bins can in fact sometimes 
go negative after background subtraction. The widely varying intensity 
falls within an envelope where the upper and lower edges are defined 
by the maxima and minima respectively of the rotation modulation 
(see Figs. 4 \& 5). In the lower panel in Fig. 1 we show the mean 
count rate averaged over complete rotation cycles according to our 
ephemeris as defined in section 3.1. Rotation cycles that are, 
for any reason incomplete are omitted, leaving a total of 344 plotted 
points. There is still a factor of $\sim$2 variation over timescales 
corresponding to tens of rotation cycles and the variability roughly 
scales with the intensity. We have fitted a Gaussian curve to these 
data points since the distribution appears to be symmetrical about the 
minimum. The center of the extended minimum is at UT 17:43:45 
$\pm$ 50 seconds on 20th July 1996 with a $\sigma$ width of 19467 
$\pm$ 78 seconds. The initial level is not well defined but for the 
fit shown is 56.4 $\pm$ 0.1 counts s$^{-1}$ with a minimum during 
interval 2 of 5.65 $\pm$ 0.2 counts s$^{-1}$. There is a suggestion of 
less variability on the climb out of the extended minima than during 
the entry into it.

\subsection{PERIOD DETERMINATION}
The most persistent feature (see Fig. 4) of the rotation modulation is 
a sharp "dip" with a phase width of $\sim0.05$. This can be 
identified in published profiles from many previous measurements and is 
evident in the mean pulse profile even at the lowest count rates during 
interval 2. We use it to define phase zero for the pulse cycle. The PCA 
data on GX 1+4 have sufficient sensitivity that even with only three  
PCUs operating most of the individual rotation dips can be seen 
throughout the observation except during the faintest part of the 
extended minimum. There is no difficulty in maintaining the cycle count 
across the many gaps evident in Fig. 1. Denoting the first dip observed as 
number 1, the last one seen is number 975. Of this set only 405 occurred 
during RXTE on source  time. The relative distributions of the data in the 3 
intervals in Fig. 1 can be summarised as follows. Interval 1 spans dips 
1 - 632 with 259 of these being during on source time and potentially 
observable, interval 2 spans dips 633 - 771 with a total of 44 being 
potentially observable and interval 3 spans dips 772 - 975 with 102 being 
potentially observable. To preserve sufficient signal to noise for the 
dip profiles, the data were first binned up into 1 second samples 
for the energy range 2 - 20 keV. We then used an initial period estimate 
to define the expected positions of the centers of the 405 dips. 
Fits were then performed to a small data window of $\pm$5 bins centered 
at each expected dip position to produce a plot of the observed 
dip time minus the calculated time (O - C), assuming a constant period, 
against dip cycle number (N). Since the individual dips can be quite 
noisy, are variable in profile, may possibly move around slightly in 
time and also do not descend from or recover to a well defined intensity level, 
we have chosen to use a simple parabolic curve for the fitting function. 
The intention is to provide a consistent estimate of the time of minimum 
count rate for each individual dip. This restricted goal also 
prompted us to fit the dips in the 2 - 20 keV light curves without the 
background subtraction applied. This has the advantage that the error 
treatment in the curve fitting process is more valid since the count rates 
are much closer to a normal distribution without reaching low, poisson 
dominated values, at the bottom of each dip. The dip total was reduced 
to only 309 after rejection of poorly fitted dips and also excluding 
all dips within interval 2 where the mean count rate was very low and the 
dips hardly detectable. We propose in a later section that these dips 
may be caused by eclipses of the hot spot by the accretion column.

A plot of the O - C residuals against dip cycle number, as in Fig. 2, 
shows considerable scatter around the mean but repeated trials, based 
on the assumption that the period is constant, allow adjustment of the period 
to provide the best fit. The linear fit for the period gives a value of 
P = 124.36568 $\pm$ 0.00020 seconds. This fit is represented by the 
horizontal dotted line across the center of Fig. 2. Since there is a 
reasonable expectation of seeing a small change in $\dot P/P$ over the 34 
hour span of the observations we have also fitted a 2nd order polynomial 
to the data which is shown by the solid curved trace in Fig. 2. The times 
of dip minima for this fit are given by 
T = [ N $\times$ 124.36213 $\pm$ 0.00080 ] 
+ [ $N^{2} \times$ 0.00000344 $\pm$ 0.00000075 ] seconds where N is the 
cycle count starting from zero at the first observed dip. These fits 
suggest that the period is not constant during the observation and 
indicate a value for $\dot P/P$ of $\sim -1.5$\% per year. The departure 
from a constant period has a 4.5$\sigma$ significance. The larger 
error bars tend to dominate in Fig. 2 but $\sim$40\% of the 309 points 
have a standard deviation of $\le$ $\pm$ 0.5 seconds. Because 
individual dip profiles are often well fitted by a minimum which 
is substantially displaced in time from the expected position (up to a 
few seconds) we have examined the series of O - C residuals for 
any periodic component. The data have many gaps and $\sim$66\% are missing 
so, after rejection of the poorly fitted dips, we have used the method 
described by Bopp et al. (1970). This has revealed no periodic trends 
in the O - C residuals so we conclude that the pulse dips move 
about, through a small range, in a random fashion. The distribution of 
the O - C residuals is approximately gaussian with a $\sigma$ width 
of $\sim$2.3 seconds which corresponds to $\sim7$ degrees in 
rotation phase. We have also compared the results obtained by repeating 
the above analysis for $\pm$4, $\pm$6 \& $\pm$7 bins in addition to the 
presented case of $\pm$5 bins. The smallest of these windows is rather 
short compared to the width of the dip features evident in Figs. 4 \& 5 
but for all these cases the inferred $\dot P/P$ values and errors are 
similar. Equivalent plots to Fig. 2 for these other cases also look 
similar, however, the individual O - C values are different in each case, 
though the spread is within the typical error bars. A plot of the O - C 
values for the $\pm$5 bin case against those for the $\pm$6 bin case has 
a correlation coefficient of 0.89 for 286 points. The scatter along 
the expected +1 slope has a $\sigma$ of $\pm$0.4 seconds. 

Historically GX 1+4 has spent most of its time in a spin down state 
(see Figure 1 \& 2, Chakrabarty et al. 1997). The {\it Burst 
and Transient Source Experiment} (BATSE) data points presented by 
Chakrabarty et al. (1997) for the long term period changes in GX 1+4 
were derived over 5 day intervals but BATSE could not detect GX 1+4 
for an extended period encompassing our observation due to its weak state. 
The BATSE data for this period, derived from 4 - 8 day intervals, are 
illustrated in Fig. 3 which shows the pulse period and $\dot P$ changes 
over a $\sim100$ day interval. The first BATSE detections of GX 1+4 after 
our observation show that spin up commenced within 10 days of our 
observation and lasted for 15 - 20 days before spin down resumed.

\subsection{PHASE FOLDED LIGHT CURVES}
The pulsar period derived in the previous section was used to construct 
phase folded pulse profiles for the entire data set in several energy 
channels. For this analysis the period was assumed to have a constant value 
of 124.36568 seconds. These pulse profiles are presented in two energy 
ranges in Figs. 4 \& 5, each of which has a curve for the intervals 
marked as 1, 2 \& 3 in Fig. 1. The pulse profiles clearly changed 
substantially during the observations. The profile during the brightest 
state (interval 1) was similar to others measured during the 1980-90's 
low state with the leading-edge brightest. Pulsations almost disappeared 
(ignoring the sharp dips) during the lowest intensity state 
(interval 2) but were again observed strongly when the flux increased once 
more (interval 3). The pulse profile however had changed significantly since 
interval 1, and was similar to those profiles measured during the 1970s, 
with the trailing-edge brightest. This change mirrors that which occurred 
between the 1970s and 1980s (Greenhill et al. 1998) but over a very much 
shorter timescale and with the change in the opposite direction. We are 
not aware of any previous observation of similar pulse profile changes 
on such short timescales.

In Table 1 we present the results from fitting gaussian profiles about 
phase 0.0 to the six dips shown in Figs. 4 \& 5. Defining the pre-dip 
level for the gaussian is somewhat problematic but we have chosen the 
'shoulder' in the count rate just prior to each dip. One could also 
argue that the dips have a rather flat bottom during the extended 
minimum through interval 2 but the count rate at the center of the dip 
is very low and the statistics relatively poor. Within the errors all 
six dips occur at exactly the same phase. The principal feature in 
Table 1 is that the dips are much shorter in duration in the 7 - 20 
keV band than in the 2 - 7 keV band. In Figs. 4 \& 5 the three 
curves have been vertically displaced for clarity but it is clear that the 
bottom of the dips, for all intervals in both figures, are almost 
coincident in count rate. A second feature in Table 1 is that the dips 
are narrower in interval 2 than in intervals 1 or 3. This suggests that 
when the source is brighter the enhanced emission is observed at all 
phases except the center of the dip. It is possible, given the 
uncertainties of the PCA background subtraction process, that the 
flat bottom of the dips for interval 2 in Figs. 4 \& 5 represents 
zero flux from GX 1+4.

\subsection{FLARE}
A significant flare is visible in Fig. 1 near the center of the extended 
minimum. The flare center is at 17:23:38 UT on 20th July 1996 which 
corresponds to a phase of 0.67 on our rotation ephemeris, defined so 
that phase 0.0 is the phase of the primary minimum. A phase of 0.67 is 
similar to the phase of the brightest part of the pulse when the 
intensity rose again in interval 3 and the pulses were seen to be 
trailing-edge bright. The flare is shown in more detail in Fig. 6. 
A smaller enhancement in emission is seen at 17:22:46 UT and may represent 
the leading part of a trailing-edge bright pulse profile. The main part 
of the flare has a duration of $\sim$6 seconds, although it is sharply 
peaked, and is superimposed on a longer somewhat triangular profile 
enhancement that covers about half of the basic $\sim$124 second rotation 
period. The flare is roughly symmetrical with no sign of the characteristic 
sharp rise and exponential decline shown by type I X-ray bursts from 
neutron stars. The flare occurs only 1207 seconds before the center 
of the extended broad minimum. The close proximity is remarkable given 
the width of the feature of $>40,000$ seconds but this may be a coincidence. 
In the 2 - 20 keV energy range the peak count rate of the flare was 
$\sim$98.6 counts s$^{-1}$ although the running average shown in Fig. 6 
reduces this to $\sim$83.9 counts s$^{-1}$. The peak rate was not exceeded 
elsewhere during a window of $\pm$6.25 hours centered on the event. 

There is a possibility that the flare originates from another X-ray source 
in the PCA field of view and not from GX 1+4 itself. The five PCA detectors 
are not quite co-aligned and for sufficiently bright flares an estimate 
can be made of their position relative to the spacecraft pointing 
direction (see 4.2, Strohmayer et al. 1997). Unfortunately this is not 
possible in this case since PCU's 1, 2 \& 3 are closely aligned and the 
count rate from the flare is too low to get meaningful results by this method. 

In Fig. 6 there is also a strong indication of a precursor mini flare 
in the preceding rotation cycle to that of the main flare. The time interval 
between the precursor mini flare and the main flare is estimated to be 
140 - 150 seconds, which is significantly longer than the neutron star 
rotation period of ~124 seconds. However, if the two enhancements at 
17:22:46 UT and 17:23:38 UT are part of the same pulse profile, then the 
separation between the two flaring episodes is similar to the pulse 
period. Galloway et al. (1999) have suggested an alternative explanation 
with these flares being due to episodes of accretion resulting from 
successive orbits of a locally dense patch of matter in the accretion 
disk. The pre-flare structure seen here in GX 1+4 is somewhat reminiscent 
of some of the bursting activity seen in GRO J1744-28 (Giles et al. 1996).

\section{DISCUSSION}
The low X-ray intensity during our 1996 observation has uncovered several 
unexpected new features of GX 1+4. The mechanism behind the pulse profile 
change may be related to the cause of the much longer term changes 
observed between the high state of the 1970s and the lower state which 
has persisted until the present day (Greenhill et al. 1998). That a similar 
pulse profile change can take place on such short timescales may provide 
information on the underlying disk dynamics and the spinup-spindown 
behaviour of the system (Greenhill et al. 1999). The origin of the very 
sharp dips in Figs. 4 \& 5 is somewhat enigmatic, partly because 
spectral changes over the observation make model fitting very difficult 
(Galloway 1999). The width of this dip feature decreases with increasing 
energy. A similar trend of decreasing width with increasing energy is 
apparent (but not remarked upon) in Ginga data (Makishima et al. 1988). 
The dip feature is present in our data during all three intervals, even 
when the flux drops to its lowest level. 

In general, models predicting pulse profiles in X-ray pulsars (Leahy \& 
Li 1995; M\'esz\'aros \& Nagel 1985) appear unable to reproduce such narrow 
dip features. Since the sharp dips are present in the pulse profiles for 
all energies up to $\sim100$ keV (White et al. 1983; Greenhill et al. 1998) 
the mechanism responsible for the dip must be effective over a very wide 
energy range. Similar sharp dips observed in profiles from other pulsars 
(Cemeljic \& Bulik 1998; Reig \& Roche 1999) have been attributed to 
eclipses of the emission region by the accretion stream. The optical 
depth for non-resonant Compton scattering, which is likely to be an important 
process in the column, will vary depending on the relative orientation of 
the column with respect to the observer. In particular the optical depth 
will reach a maximum at the closest approach of the line of sight to the 
magnetic field axis because the line of sight then passes through a larger 
slice of the accretion column. The additional scattering for this 
alignment will produce a corresponding dip in the pulse profile which 
can be moderately sharp over a range of different geometries 
(Galloway, 1999). Another possible cause of a sharp dip is resonant 
cyclotron absorption which has been discussed by Dotani et al. (1989) 
and Greenhill et al. (1993). For this absorption process the emission at 
frequency $\nu$ can be very efficiently absorbed by the accretion plasma 
when the local cyclotron energy in the accretion stream is the same as 
$\nu$. The accretion column becomes wider with height above the polar cap 
as the field lines diverge and the frequency that is most effectively 
absorbed also decreases with height as the magnetic field decreases. 
The scattering region in the column, at a height corresponding to each 
particular frequency, is expected to extend over a greater area than the 
polar cap below it and so can completely cover this X-ray "hot spot" for a 
substantial range of viewing angles. Hence, the line of sight does not 
need to be very closely aligned with the magnetic field axis for a 
significant dip to be created. Additionally, this model predicts that the 
width of the dip minimum should decrease with increasing energy as is 
seen in the data presented in this paper. It seems feasible that both the 
processes described above may be operating simultaneously in producing 
the observed dip. A further, but less likely, possibility (Storey et al. 
1998; Greenhill et al. 1998) is that the line of sight is closest to the 
magnetic axis at phase 0.5 and that the primary dip represents the edge 
of a very broad pulse centered at phase 0.5. The asymmetry in the pulse 
profile could then be caused by an asymmetry in the accretion flow onto 
the polar cap region. A model of this type does not however provide a 
simple explanation for the energy dependence of the dip width. 

The results presented above show that we have observed GX 1+4 during a 
transition from leading-edge bright pulses to trailing-edge bright 
pulses. There is strong evidence that this was associated with a 
reversal of torque in GX 1+4 and possibly a change in accretion disk 
spin direction. This observation and the modelling of Greenhill et al. 
(1999) suggests that our understanding of the processes of torque transfer 
in accreting X-ray pulsars would greatly benefit from more detailed 
observations of GX 1+4 through a torque reversal period. The flare 
observed during the time of intensity minimum may also provide new 
insights into the mechanisms of mass transfer from the accretion disk to 
the neutron star if more examples can be observed. There is marginal 
evidence for wandering in the phases of the individual sharp dips but the 
count rates are too low to allow detailed study of any such effect. 
If present, this could be interpreted as evidence for wander in the 
position of the accretion column if the dips are due to absorbtion 
by accreting matter. The reality of this hypothetical wandering could be 
tested by analysing observations of GX 1+4 obtained while it was in a 
higher intensity state.

\acknowledgments 
We thank the RXTE Science Operations Center planning and operations 
staff for successfully conducting these observations. We also thank 
Kinwah Wu for helpful advice, Jean Swank for support and useful 
discussions and Keith Jahoda for comments on the PCA background.

\pagebreak
\begin{table*}
\begin{center}{Table 1: Averaged profile parameters for the dip features in Figs. 4 \& 5}
\begin{tabular}{clll} \\
\tableline
\tableline
Interval  &  Parameter     &  2 - 7 keV          &  7 - 20 keV \\
\tableline
1  &  Center (phase)       &  0.9997 $\pm$ 0.0011  &   1.0002 $\pm$ 0.0008  \\
   &  Half Width (sec)     &  3.82   $\pm$ 0.18    &   2.78   $\pm$ 0.12    \\
   &  Depth (c s$^{-1}$)   &  16.07  $\pm$ 0.57    &   18.61  $\pm$ 0.63    \\
   &  Minima (c s$^{-1}$)  &  2.04   $\pm$ 0.69    &   3.87   $\pm$ 0.71    \\
\tableline
2  &  Center (phase)       &  1.0030 $\pm$ 0.0029  &   0.9996 $\pm$ 0.0018  \\
   &  Half Width (sec)     &  2.49   $\pm$ 0.40    &   2.02   $\pm$ 0.25    \\
   &  Depth (c s$^{-1}$)   &  4.59   $\pm$ 0.61    &   6.42   $\pm$ 0.68    \\
   &  Minima (c s$^{-1}$)  & -0.58   $\pm$ 0.65    &  -0.03   $\pm$ 0.72    \\
\tableline
3  &  Center (phase)       &  1.0036 $\pm$ 0.0020  &   1.0026 $\pm$ 0.0011  \\
   &  Half Width (sec)     &  4.24   $\pm$ 0.35    &   2.77   $\pm$ 0.16    \\
   &  Depth (c s$^{-1}$)   &  8.80   $\pm$ 0.52    &   12.74  $\pm$ 0.58    \\
   &  Minima (c s$^{-1}$)  &  1.69   $\pm$ 0.64    &   3.29   $\pm$ 0.64    \\
\tableline
\end{tabular}
\end{center}
\end{table*}

\pagebreak
\centerline{\bf Figure Captions}

{\bf  Figure 1.}  The light curve of the $\sim$34 hour observation of 
GX 1+4. The count rate shown is a running 4 second average, with a one 
second resolution, for the full energy range of the PCA (2 - 20 keV). 
The rapid intensity variation reflects the pulse amplitude of the 
$\sim$124.4 second rotation period and this is superimposed on a broad 
dip in intensity down to a minimum level of $\leq0.7$ mcrab. The flare is 
visible near the center of this broad minimum. The lower panel plots 
mean count rates for individual complete rotation cycles and a Gaussian 
fit through these data points. 

{\bf  Figure 2.}  This plot shows the trend in the fits to the dip 
minima for the 974 cycles covered by the observation in Fig. 1. 
Located and fitted individual dip centers are plotted plus or minus with 
respect to an ephemeris period of 124.36568 seconds. The error bars on 
the O - C residuals are $\pm$1$\sigma$. The horizontal dotted line is 
the best linear fit and the solid curved line is the best fit 2nd order 
polynomial. The lower dotted curve is for a $\dot P/P$ of 1.5\% per year.

{\bf  Figure 3.}  The BATSE data for GX1+4 over the period from 
20 April 1996 to 25 August 1996. The RXTE observation, at HJD 
$\sim2450285$, is marked with the small square and occurs just 
before a brief interval of spin up commenced. The $\dot P$ axis in 
the lower part of the figure is in units of 10$^{-8}$ s s$^{-1}$. 

{\bf  Figure 4.}  Phase folded light curves for the energy range 2 - 7 keV. 
The three traces correspond to intervals 1, 2 \& 3 in Fig. 1. 
To differentiate the three profiles trace 1 has been offset vertically 
by 10.0 counts s$^{-1}$ and trace 3 by 5.0 counts s$^{-1}$. The large 
rotation modulation has shifted character between traces 1 \& 3 with 
the maximum flux occurring in the earlier part of the cycle for 
trace 1 and the latter part for trace 3.

{\bf  Figure 5.}  Phase folded light curves similar to those in Fig. 4 but 
for the higher energy range of 7 - 20 keV. Trace 1 has again been offset 
vertically by 10.0 counts s$^{-1}$ and trace 3 by 5.0 counts s$^{-1}$. The 
change in character for traces 1 \& 3 noted for Fig. 4 is repeated for this 
energy band. Fits to the dip profiles in Figs. 4 \& 5 are given in Table 1
and the feature is clearly sharper at higher energies.  

{\bf  Figure 6.}  The flare seen during the extended minima of GX1+4. 
To emphasize this feature the 2 - 20 keV count rate is plotted as a 
running 4 second average with one second resolution. The vertical dotted 
lines mark the times of the predicted dip minima.

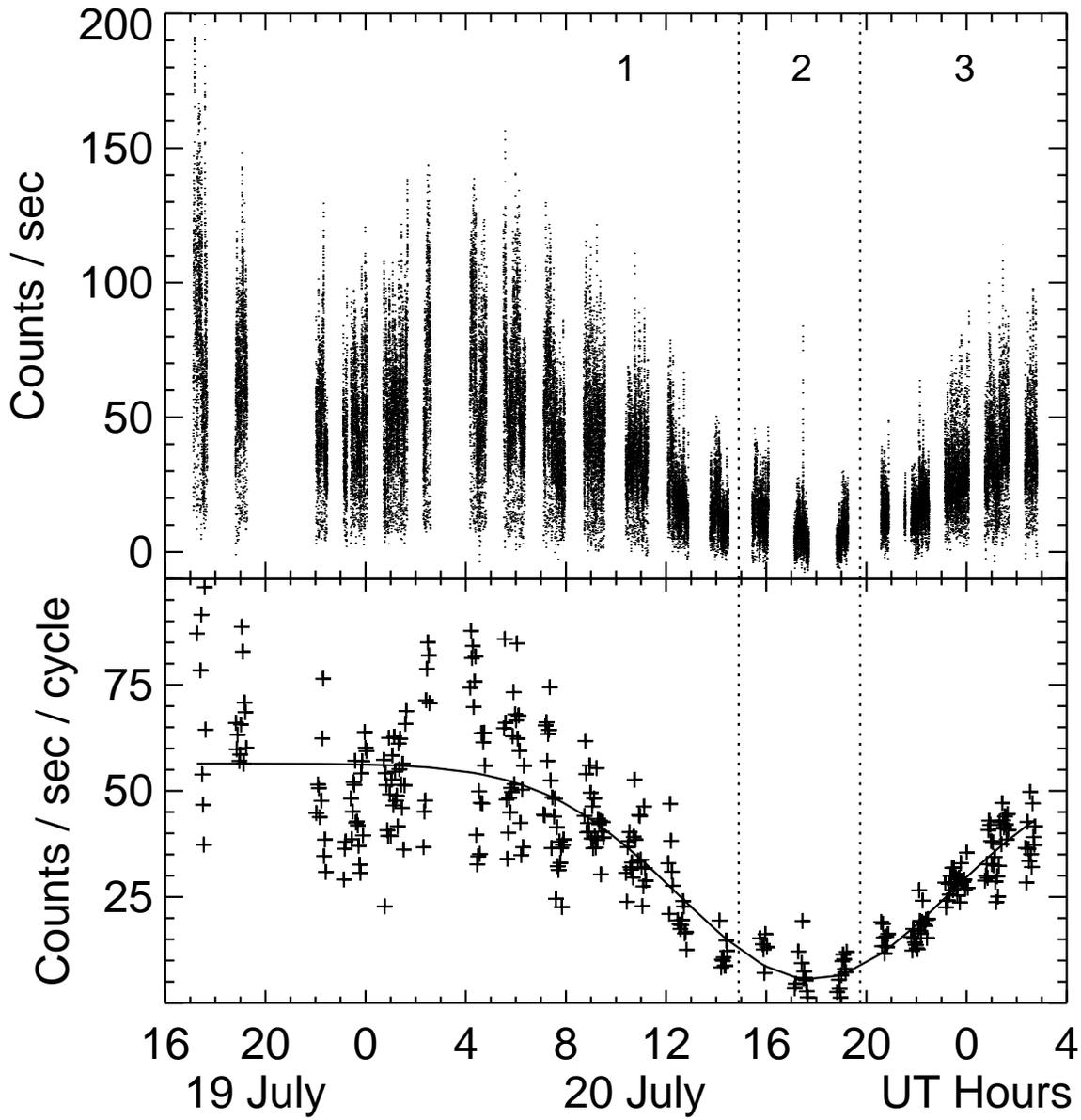
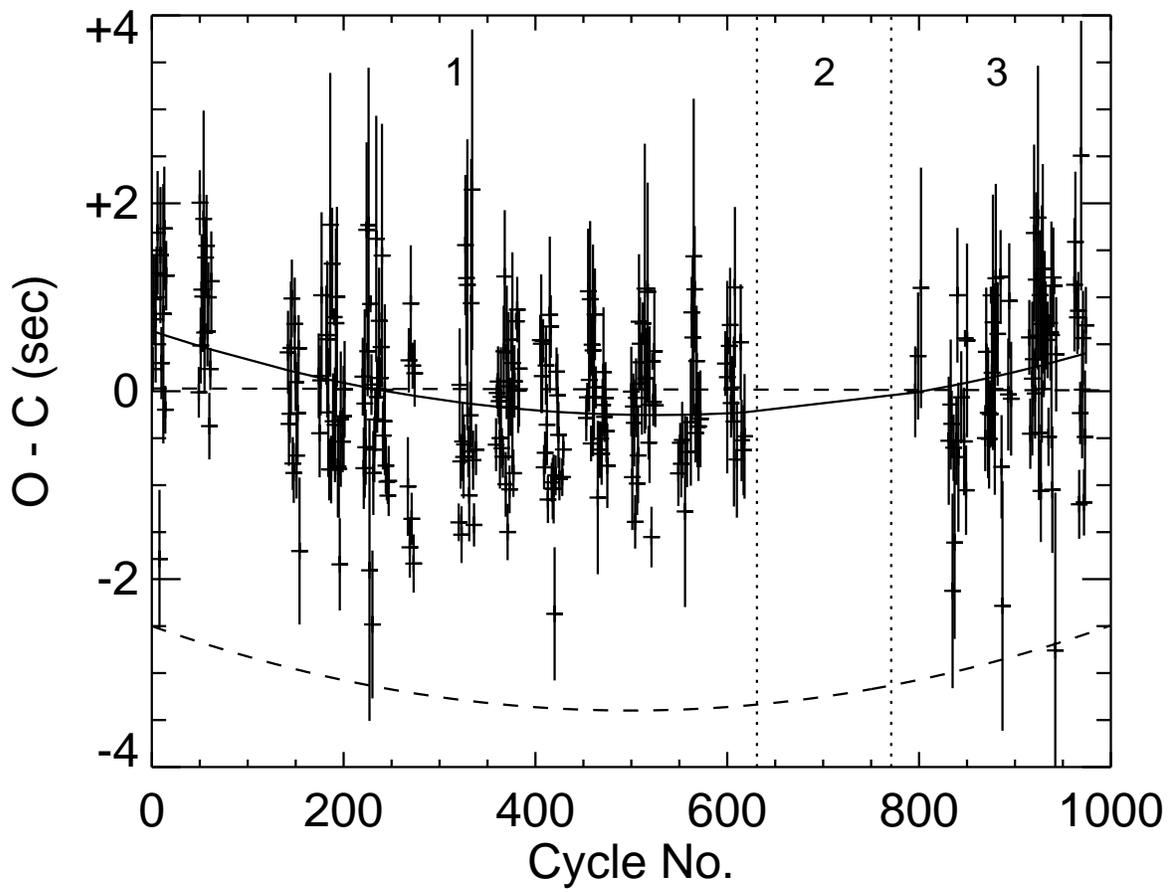
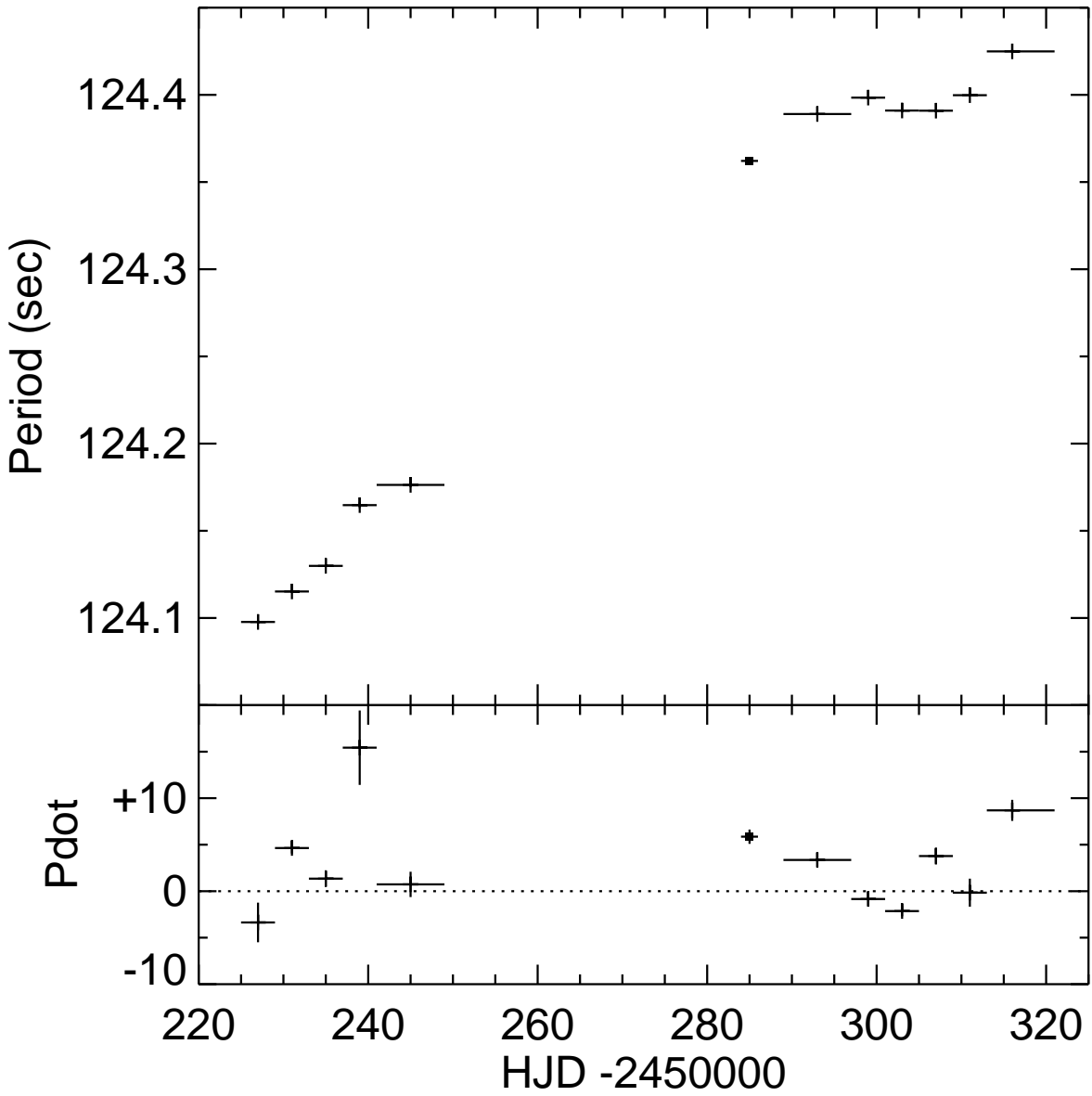
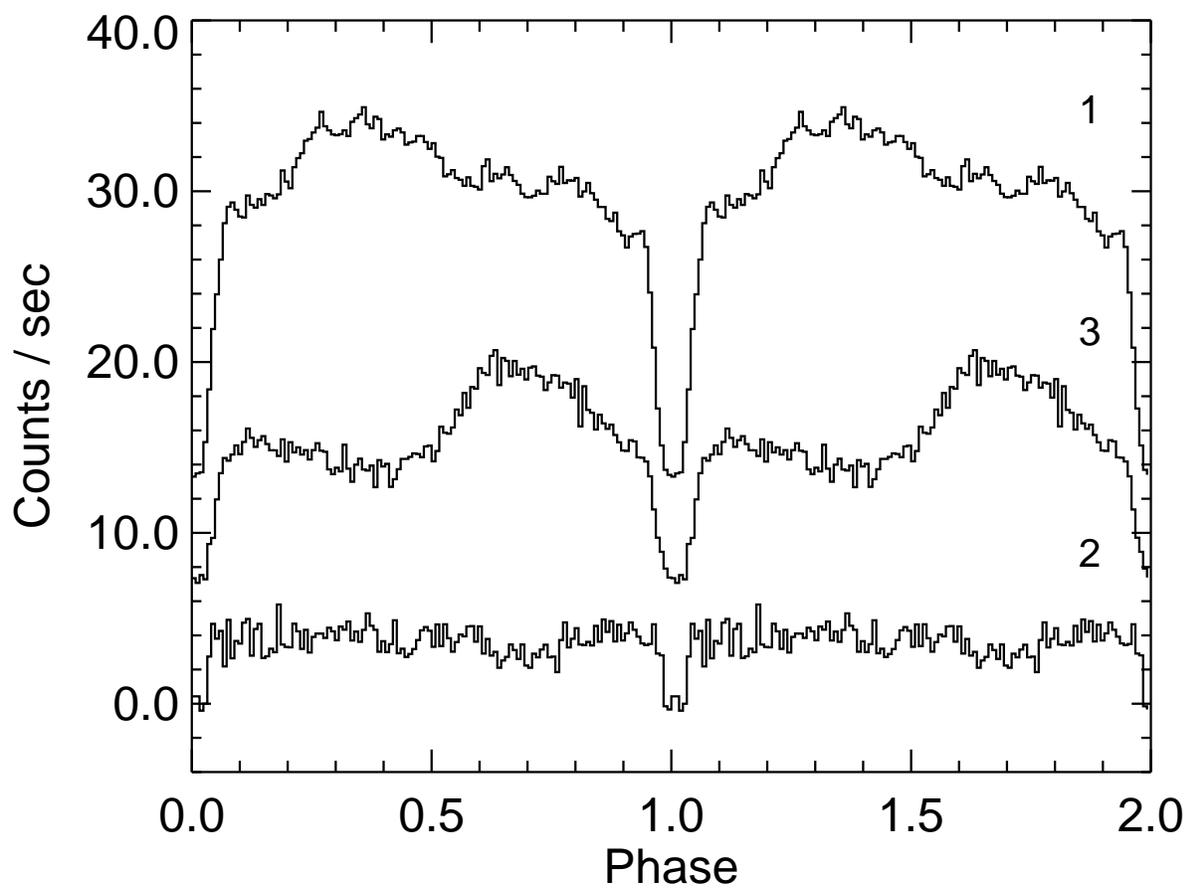
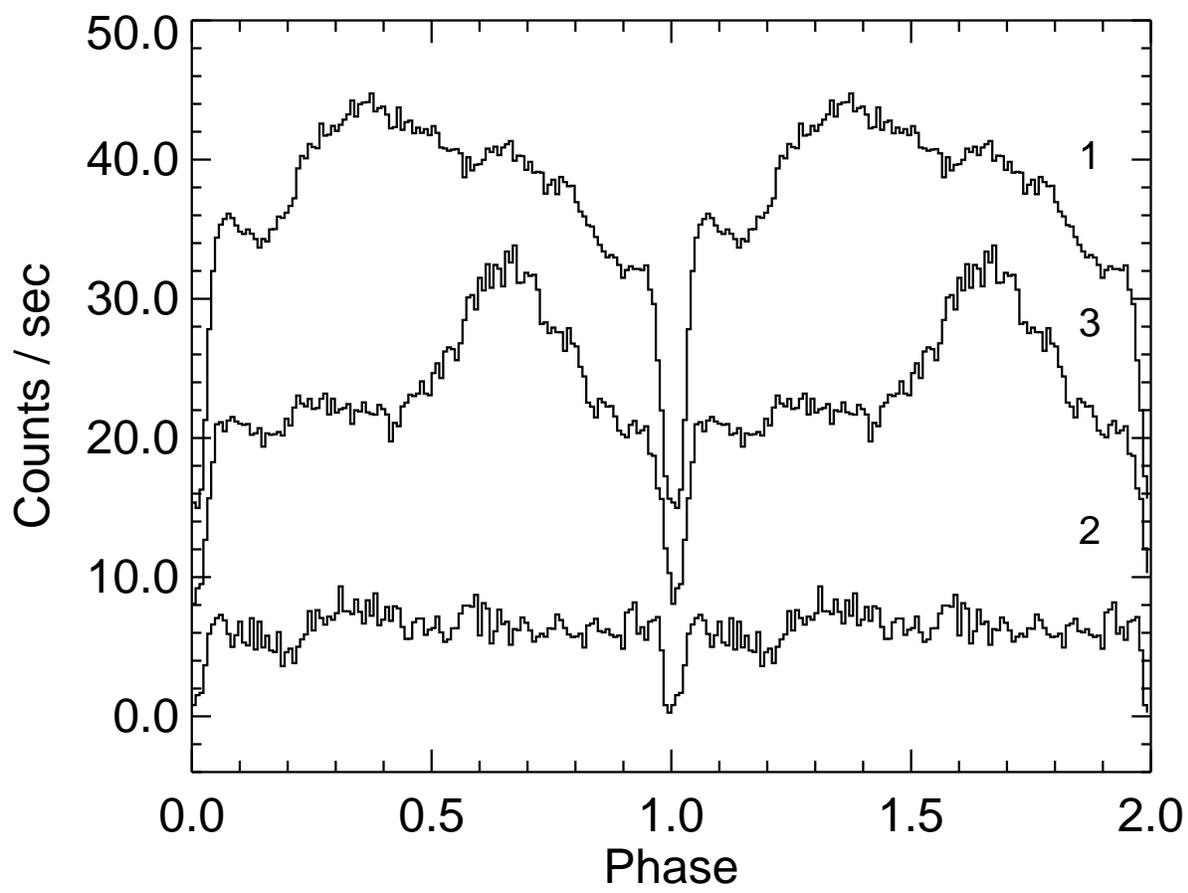
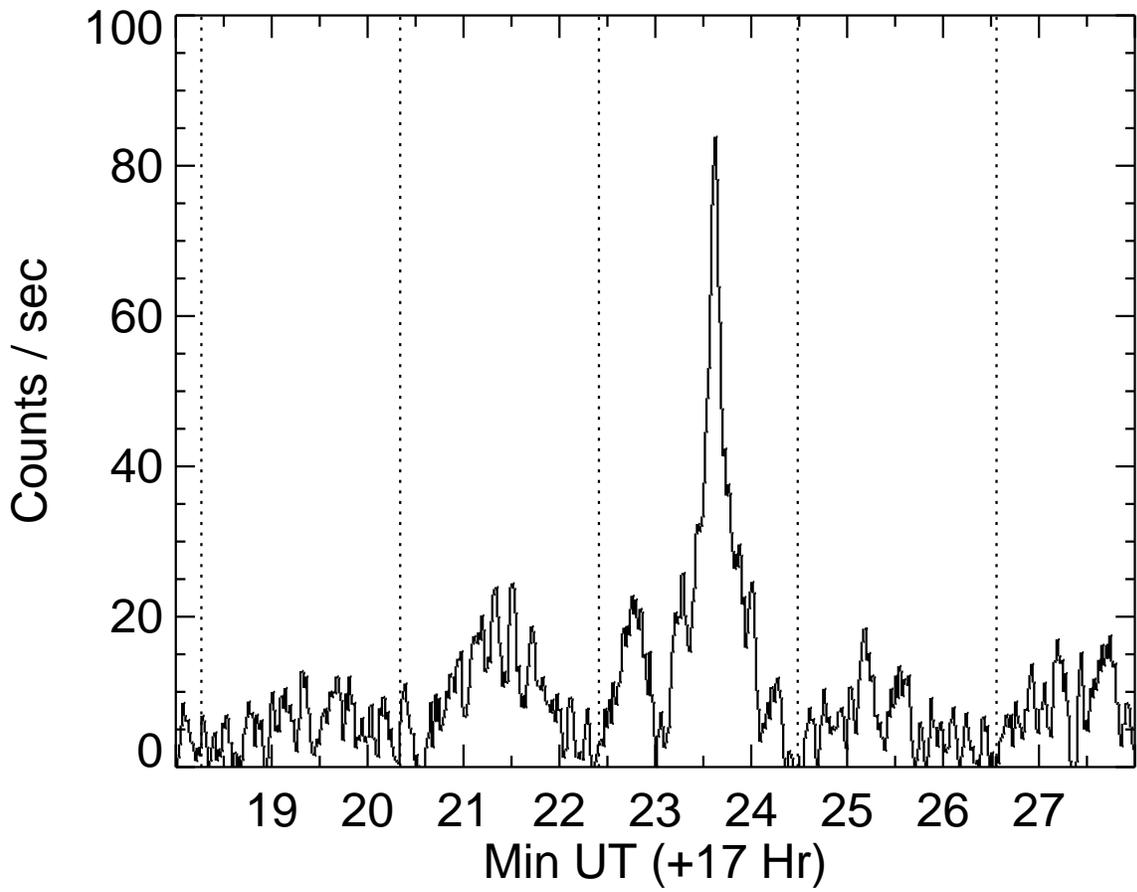
\end{document}